\documentclass[a4paper,10pt]{article}
\usepackage[dvips]{graphicx}
\usepackage{amsfonts,amsmath,amssymb}
\begin{document}
\title{Casimir Energy for a Coupled Fermion-Soliton System}

\author{L. Shahkarami\footnote{Electronic address: l$\_$shahkarami@sbu.ac.ir}, A. Mohammadi\footnote{Electronic address: a$\_$mohammadi@sbu.ac.ir} and S.S. Gousheh\footnote{Electronic address: ss-gousheh@sbu.ac.ir}\\
\small  Department of Physics, Shahid Beheshti University G.C., Evin, Tehran
19839, Iran}

\maketitle
\begin{abstract} In this paper we compute the Casimir energy for a coupled fermion-pseudoscalar field system.
 In the model considered in this paper the pseudoscalar field is \textit{static} and \textit{prescribed} with two adjustable parameters.
  These parameters determine the values of the field at infinity ($\pm \theta_0$) and its scale of variation ($\mu$).
    One can build up a field configuration with arbitrary topological charge by changing $\theta_0$, and interpolate between the extreme adiabatic and non-adiabatic regimes
    by changing $\mu$.
     This system is exactly solvable and therefore we compute the Casimir energy exactly and
     unambiguously by using an energy density subtraction scheme.
       We show that in general the Casimir energy goes to zero in the extreme adiabatic limit, and in the extreme non-adiabatic limit when the asymptotic
        values of the pseudoscalar field properly correspond to a configuration with an arbitrary topological charge. Moreover,
         in general the Casimir energy is always positive and on the average an increasing function of $\theta_0$ and always has local maxima when there is a zero mode,
          showing that these configurations are energetically unfavorable. We also compute and display the energy densities associated with the spectral deficiencies in both of the continua, and those of the bound states. We show that the energy densities associated with the distortion of the spectrum of the states with $E>0$ and $E<0$ are mirror images of each other. We also compute and display the Casimir energy density. Finally we compute the energy of a system consisting of a soliton and a valance electron and show that the Casimir energy of the system is comparable with the binding energy.
\end{abstract}
\section{Introduction}
Dutch physicist Hendrick Brugt Gerhard Casimir published his famous paper \cite{casimir} in 1948, where he found a simple yet profound explanation for the retarded
 van der Waals interaction between polarizable molecules. He showed that there exists a net force between grounded infinite metallic plates, placed a few micrometers
  apart,
  in a vacuum without any external electromagnetic field. The Casimir effect is essentially a direct consequence of the change in the vacuum energy of a quantum field when
  some boundary condition or background field is imposed. Since the Casimir's work, many papers have been written on the Casimir energy or the resulting force due to
  the presence of boundary conditions. They have calculated this effect for
  parallel plates \cite{plate}, cylinders \cite{cylinder}, and
 spheres \cite{sphere} (first considered by Casimir \cite{casimir2}, DeRaad and Milton \cite{deraad}, and Boyer \cite{boyer}, respectively), and many other geometries \cite{othergeometry}. In addition, many authors have considered this effect for various fields, different boundary conditions, dimensions and topologies \cite{dim}.
  There are various methods to calculate the Casimir energy, e.g. the Green function formalism \cite{green}, the heat-kernel method \cite{heat} and
   the multiple-scattering expansions \cite{scatter}. Also, as mentioned before, the presence of the nontrivial background fields such as solitons affect
    the zero-point energy. For example, to compute the quantum corrections to the mass of the soliton, the change in the quantum fluctuations of the vacuum energy
     caused by the presence of the soliton, should be taken into account. In 1974, Dashen \textit{et al} \cite{dashen} computed the one-loop correction to the mass
      of the kink of the $\phi^4$ theory for the first time. After this work, several authors have used various methods to compute the corrections to the soliton mass,
       such as the scattering phase shift method, zeta function analytic continuation technique, dimensional regularization technique, etc. \cite{solitonmass,heavy}.
        Moreover, the Casimir effect appears in supersymmetric models. To investigate the validity of the BPS saturation by supersymmetric solitons, one should
        compute the change in the fluctuations of the vacuum energy to obtain the quantum corrections to the soliton mass (see for example \cite{super}).
\par One of the first experimental attempts to verify the existence of the Casimir effect was conducted by Marcus Sparnaay for two parallel metallic plates
 in 1958 \cite{sparnaay}. The results was not in contradiction with the Casimir theory, but had a very poor accuracy.
  In 1997, using a plate and a spherical lens covered by a metallic sheet,
  Steve K. Lamoreaux measured the Casimir energy with high accuracy \cite{Lamorea}. It was the first ``successful'' experiment to verify the Casimir effect.
   Since then, many different experiments have been performed to measure the casimir effects for various geometries, such as two parallel plates and a sphere
    in front of a plane, with ever increasing accuracy (see for example \cite{exper}).
\par In this paper we consider a pseudoscalar field as a prescribed, spatially varying background field which is coupled chirally to a Fermi field.
 The background field in our model has two adjustable parameters, one related to its scale of variation and the second to its asymptotic value at spatial infinity.
  By varying the latter from zero to $n\pi$, we can build a field configuration with topological charge $n$. As it is well-known
  the nomenclature ``soliton'' refers to a static field configuration which is a solution to the classical field equations with finite energy and localized energy density. Topological solitons by definition carry nonzero topological charge. In a field theory in which this background field interacts with a fermion, such as the one considered in this paper, the exact shape of the soliton depends on the back-reaction of the Fermi field, which in turn depends on the particular fermionic state to which the soliton is coupled. Therefore, the soliton can in principle acquire infinitely many different shapes. However, since the back-reaction is usually small, all these different shapes are close to each other. However, we have obtained the exact (to within numerical approximation) non-perturbative solutions for the coupled system treating both fields as dynamical, verifying the perturbative result that the back-reaction is small \cite{accepted}. The problem of obtaining the Casimir energy for any of these particular field configurations is also not exactly solvable. Therefore, in this paper we choose a simple field configuration which has proper topological charge and renders the problem exactly solvable. Although the field configuration that we choose is not a solution to the field equations, it possesses all other properties of a proper soliton. We shall henceforth refer to this configuration as a soliton for brevity of notation. We expect the exact Casimir energy for this topologically non-trivial configuration to illuminate many of the qualitative features of the problem with an exact soliton. Since this model is exactly solvable and we have the
   complete spectrum of the Fermi field \cite{gousheh}, we are able to calculate the vacuum expectation value of the Hamiltonian directly and exactly
    for all values of the aforementioned parameters. We obtain the Casimir energy for this system by subtracting the zero-point energy in the presence and
     absence of the background field. In Section II we introduce the model and present a derivation for the expression of the Casimir energy. In Section III we obtain
       an exact expression for the Casimir energy and its density for our model. Then, we thoroughly explore the properties and behavior of
        the Casimir energy  as a function of the parameters of our model. First, we depict the Casimir energy as a function of the scale of variation
         of the soliton with winding number one. In general there is a sharp maximum whenever the Fermi field has a zero-mode bound state.
          We obtain the expected results that the casimir energy tends to zero in the limits of extremely wide and sharp solitons. Second, we depict the Casimir energy as a function of $\theta_0$. In general $E_{\text{Casimir}}$ is an increasing function of $\theta_0$ with very mild maxima in the form of cusps whenever there is an energy level crossing zero.
           Then, we exhibit the energy densities of the bound and continuum states and their sum for states with $E<0$ and $E>0$, separately. These two energy densities are exact mirror images of each other, and from this we compute the Casimir energy density.
             In Section IV we summarize and discuss the results.
\section{A fermion-pseudoscalar field model and the Casimir energy}
\begin{center}
\begin{figure}[th] \hspace{3.5cm} \includegraphics[width=6.3cm]{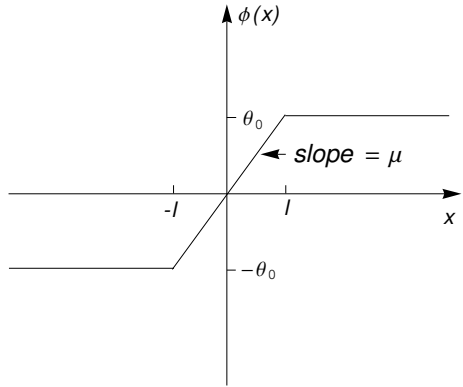}\caption{\label{fig.1} \small
  The form of the background pseudoscalar field $\phi\left(x\right)$.}
  \label{geometry}
\end{figure}
\end{center}

Consider the following Lagrangian describing a fermion interacting
non-linearly with a pseudoscalar field $\phi\left(x\right)$, in
($1+1$) dimensions,
\begin{equation}\label{e1}\vspace{.2cm}
  {\cal L}= {\bar{\psi}}\left(x,t\right)\left(i \gamma^\mu \partial_\mu
 - m \text{e}^{i\phi\left(x\right)\gamma^5}\right)\psi \left(x,t\right).
\end{equation}
In order to have an exactly solvable model, Gousheh and Mobilia \cite{gousheh} chose $\phi\left(x\right)$
to be piecewise linear as follows
\begin{equation}\label{e2}
\phi(x)=\begin{cases}
-\theta_0& \text{for } x \leqslant -l ,\\
\mu x& \text{for }-l\leqslant x\leqslant l,\\
+\theta_0 & \text{for } l \leqslant x,
\end{cases}
\end{equation}
where $\mu=\theta_0/l$ is the slope in the middle region. One could interpolate between the extreme adiabatic
and non-adiabatic regimes by changing $\mu$ from zero to infinity, respectively. When $\theta_0=n\pi$, the soliton has winding number $n$.
 Figure (\ref{fig.1}) shows the form of $\phi\left(x\right)$. As mentioned earlier this simple topological configuration is used as an approximation to the kink of the $\lambda \phi^4$ theory. Although our simple static configuration is not a solution to the Euler-Lagrange equation, its total energy can be easily obtained from the corresponding Hamiltonian as follows

 \begin{equation}\label{e1r}\vspace{.2cm}
  M_{\text{cl}}=\int_{-\infty}^{\infty} {\cal H}[\phi(x)]dx=\int_{-\infty}^{\infty} \left[\frac{1}{2}(\partial_x \phi )^2 +\frac{\lambda}{4}\left(\phi^2-\frac{m'^2}{\lambda}\right)^2\right]dx=\frac{23}{15}\mu\theta_0,
\end{equation}
where $\frac{m'}{\sqrt{\lambda}}=\theta_0$, the value of $\phi(x)$ when $x\rightarrow\infty$, and $\frac{m'^2}{\sqrt{2\lambda}}=\mu$, the slope at $x=0$.
\begin{center}
\begin{figure}[th] \hspace{3.cm} \includegraphics[width=8.1cm]{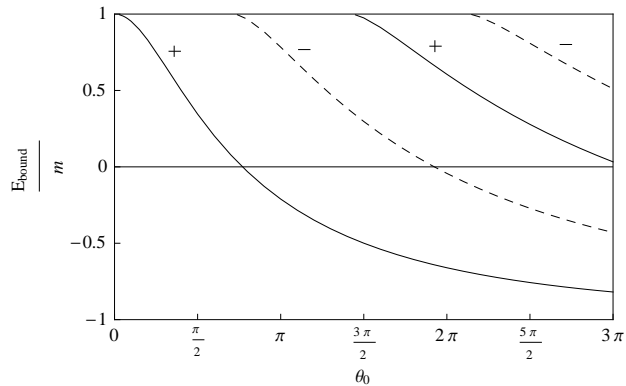}\caption{\label{fig.2} \small
  The energies of the bound states as a function of $\theta_0$ for $\mu=4m$ obtained from $\frac{m}{\lambda}\frac{\sin\left(\zeta l\right)}{\zeta l}\theta_0=\pm 1$ derived in \cite{gousheh}. Solid (dashed) lines show the bound states with positive (negative) parity. }
  \label{geometry}
\end{figure}
\end{center}
The procedure is to find the spectrum of the Fermi field in the presence of this background, for arbitrary values of $\mu$ and $\theta_0$.
 Then, for fixed $\mu$, one could observe the changes in the spectrum of the fermion as a soliton with the topological charge $n$ is built by changing
  $\theta_0$ from zero to $n\pi$. This method was first introduced by MacKenzie and Wilczek \cite{mackenzie1} and is based upon the adiabatic method first
   introduced by Goldstone and Wilczek \cite{goldstone}. Since this model is exactly solvable, the complete spectrum of the Fermi field, including the
    eigenfunctions and their corresponding eigenenergies, has been obtained exactly for arbitrary values of $\mu$ and $\theta_0$, and are given in \cite{gousheh}. Here we do not present the solutions in detail and the interested reader is referred to \cite{gousheh}.
If we denote the Fermi field in ($1+1$) dimensions by $\psi=\dbinom{\psi_1}
{\psi_2}$,
then we can define
\begin{equation}\label{e3}
\xi=e^{-iEt}\dbinom{\xi_1}{\xi_2}=\dbinom{\psi_1+i\psi_2}
{\psi_1-i\psi_2},
\end{equation}
and solve the equations of motion in terms of $\xi$. Since the Hamiltonian commutes with the parity operator,
 the eigenfunctions of the Hamiltonian can be chosen to be eigenfunctions of the parity operator as well. In the first representation ($\gamma^0=\sigma_1$, $\gamma^1=i\sigma_3$, $\gamma^5=\gamma^0\gamma^1=\sigma_2$), the parity
operation is given by $P\psi\left(x,t\right)=\sigma_1
\psi\left(-x,t\right)$. In the new representation i.e. Eq.\,(\ref{e3}) ($\gamma^0=\sigma_2$, $\gamma^1=i\sigma_1$, $\gamma^5=\gamma^0\gamma^1=\sigma_3$), it becomes $P\xi\left(x,t\right)=-\sigma_2
\xi\left(-x,t\right)$.
 Figure (\ref{fig.2}) shows the energies of the bound states as a function of $\theta_0$ for $\mu=4 m$ as an example.
   As shown in the figure, the bound states have alternating parity signs, starting with the positive sign. This graph will be important for the upcoming discussions.
\par In order to obtain the Casimir energy, we should subtract the zero point energy in the absence from the presence of the background field. Therefore,
 the vacuum energies for both cases should be calculated.
MacKenzie and Wilczek \cite{mackenzie1} assumed that the solutions in the presence of the soliton are complete. Later on Gousheh and Mobilia proved this assumption.
Therefore, the Fermi field operator can be expanded in terms of either the modes in the absence or in the presence of the background field, as follows
\begin{eqnarray}\label{e6}\vspace{.2cm}
\psi(x,t)&=&\int_{-\infty}^{+\infty}\frac{dk}{2\pi}\left[b_k u_k(x,t)+d_k^\dag v_k(x,t)\right]\nonumber\\
&=&\int_{0}^{+\infty}\frac{dp}{2\pi}\sum\limits_{j=\pm}\left[a_p^j
\mu_p^j(x,t)+c_p^{j\dag} \nu_p^j(x,t)\right]
+\sum\limits_{i}\left[e_i
\chi_{1b_i}(x,t)+f_i^\dag\chi_{2b_i}(x,t)
\right]
\end{eqnarray}
where $u_k$ ($v_k$) are the positive
(negative) energy continuum eigenstates in the absence of the background field. Also, $\mu_p^j$ ($\nu_p^j$) are the positive (negative) energy
continuum eigenstates and $\chi_{1b_i}$ ($\chi_{2b_i}$)  are the positive
(negative)  energy bound states in the presence of the soliton. Here, $\sum_i$ refers to the
possibility of multiple bound states and the superscript $j$ refers to the parity of
the continuum states. MacKenzie and Wilczek chose the coefficients of
both positive and negative energy bound states to be annihilation operators.
However, we choose annihilation (creation) operator for the positive
(negative) energy bound states, as we do for the continuum states. We shall see one of the advantages of this choice when we compute the Casimir energy.
In general the Hamiltonian can be expressed as follows
\begin{equation}\label{e9}\vspace{.2cm}
H=\int_{-\infty}^{+\infty}dx\left[\psi^\dag(x,t){\cal
H}\psi(x,t)\right].
\end{equation}
Substituting the expression for the field operator $\psi$ in Eq.\,(\ref{e6}) in the presence of the background field into Eq.\,(\ref{e9}), we obtain
\begin{eqnarray}\label{e61}\vspace{.2cm}
H&=&\int_{-\infty}^{+\infty}dx\bigg\{\int_{0}^{+\infty}
\frac{dp}{2\pi}\sum\limits_{j=\pm}\left[a_p^{j\dag}\mu_p^{j\dag}(x,t)+c_p^{j} \nu_p^{j\dag}(x,t)\right]+\sum\limits_{i}\left[e_i^\dag
\chi_{1b_i}^\dag(x,t)+f_i\chi_{2b_i}^\dag(x,t)
\right]\bigg\}\nonumber\\
&\times&\bigg\{\int_{0}^{+\infty}\frac{dq}{2\pi}\sum\limits_{n=\pm}\left[\sqrt{q^2+m^2}a_q^n
\mu_q^n(x,t)-\sqrt{q^2+m^2}c_q^{n\dag} \nu_q^n(x,t)\right]\nonumber\\
&+&\sum\limits_{l} \left[E^{l+}_{\text{bound}}e_l
\chi_{1b_l}(x,t)+E^{l-}_{\text{bound}}f_l^\dag\chi_{2b_l}(x,t)
\right] \bigg\},
\end{eqnarray}
where, the $\pm$ superscripts on $E_{\text{bound}}^{l}$ denote the sign of the bound state. Upon calculating the expectation value of the Hamiltonian in the vacuum state, all the terms vanish except the terms with an annihilation operator on the left side and the corresponding creation operator on the right side. There are only two terms in this form which can be written in the following form by the use of the anticommutation relations
\begin{equation}\label{e8}\vspace{.2cm}
c_p^j c_q^{m\dag}=-c_q^{m\dag}c_p^j+ \delta^{mj}\delta_{qp}\,\,\,,\,\,\, f_i f_l^{\dag}=-f_l^{\dag}f_i+ \delta_{il}.
\end{equation}
Therefore, we obtain the following expression for the vacuum expectation value of the Hamiltonian
\begin{eqnarray}\label{e11}\vspace{.2cm}
\left\langle\Omega\left|H\right|\Omega
\right\rangle&=&\int_{-\infty}^{+\infty}dx\int_{0}^{+\infty}\frac{dp}{2\pi}\sum\limits_{j=\pm}\left(-\sqrt{p^2+m^2}\right)\nu_p^{j\dag}(x,t)\nu_p^j(x,t)\nonumber\\
&+&\int_{-\infty}^{+\infty}dx\sum\limits_{i}\left(E_{\text{bound}}^{i-}\right)\chi_{2b_i}^\dag(x,t)\chi_{2b_i}(x,t),
\end{eqnarray}
where $|\Omega\rangle$ is the vacuum state in the presence of the background field.
 From this expression we conclude that the
 entire contribution to the vacuum energy for fermionic systems originates from the negative energy eigenstates.
  Note that our separation of the bound states into the positive and negative energy ones has the advantage of the automatic inclusion of the
   negative energy bound states in the expression for the Casimir energy. To obtain the Casimir energy,
 one should subtract the zero point energy in the absence from the presence of the background field.
  Thus, we obtain
\begin{eqnarray}\label{e12}\vspace{.2cm}
E_{\text{Casimir}}&=&\left\langle\Omega\left|H\right|\Omega
\right\rangle-\left\langle0\left|H_{\text{free}}\right|0
\right\rangle\nonumber\\
&=&\int_{-\infty}^{+\infty}dx\int_{0}^{+\infty}\frac{dp}{2\pi}\sum\limits_{j=\pm}\left(-\sqrt{p^2+m^2}\right)\nu_p^{j\dag}\nu_p^j+
\int_{-\infty}^{+\infty}dx\sum\limits_{i}\left(E_{\text{bound}}^{i-}\right)\chi_{2b_i}^\dag\chi_{2b_i}\nonumber\\
&-&\int_{-\infty}^{+\infty}dx\int_{-\infty}^{+\infty}\frac{dk}{2\pi}\left(-\sqrt{k^2+m^2}\right)v_k^\dag
v_k.
\end{eqnarray}
Note that our problem possesses charge conjugation symmetry and therefore our expression for the Casimir energy given in Eq.\,(\ref{e12}) is equivalent to the conventional one, where one would sum over all modes symmetrically with a factor of $1/2$, while preserving the sign.
\section{The Casimir energy and its density for this model}
\begin{center}
\begin{figure}[th] \hspace{3.5cm} \includegraphics[width=8.cm]{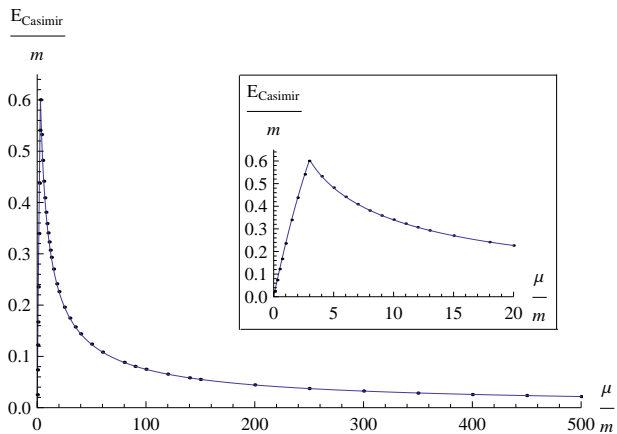}\caption{\label{fig.3} \small
  The Casimir energy as a function of $\mu$, the scale of variation of the soliton, at $\theta_0=\pi$.
   The points displayed are the results of the numerical integration and the solid line represents the best fit to the points.
    In the illustrated box we focus on small values of $\mu$ to show the details of the maximum. This maximum is precisely a cusp.}
  \label{geometry}
\end{figure}
\end{center}

By substituting the expressions for the eigenstates into Eq.\,(\ref{e12}), the Casimir energy can be calculated.
 Our prescription for subtracting two quadratically divergent
 integrals appearing in this equation,
 is to subtract the integrands with corresponding values of $p$ and $k$, and then performing the integral.
 The change in the energy densities of the Dirac sea ($E\leqslant -m$) and sky ($E\geqslant m$) (i.e. the difference between the energy densities in the presence and absence of the background field) are
\begin{eqnarray}\label{r4}\vspace{.2cm}
\varepsilon_{\text{sky,sea}}(x)=\int_{0}^{+\infty}\frac{dp}{2\pi}
\sum\limits_{j=\pm}\left(E\right)\xi_p^{j\dag}(x)\xi_p^j(x)
-\int_{-\infty}^{+\infty}\frac{dk}{2\pi}\left(E_{\text{free}}\right)\xi_k^{\text{free}\dag}(x)
\xi_k^{\text{free}}(x),
\end{eqnarray}
where $E=\pm \sqrt{p^2+m^2}$, $E_{\text{free}}=\pm \sqrt{k^2+m^2}$  ($\pm$ are for the Dirac sky and sea, respectively),
$\xi_p^j=\mu_p^j(\nu_p^j)$ for interacting Dirac sky (sea) and $\xi_k^{\text{free}}=u_k(v_k)$
for free Dirac sky (sea). The bound state energy densities are
\begin{equation}\label{r5}\vspace{.2cm}
\begin{split}
\varepsilon_{\text{bound}\pm}(x)=
\sum\limits_{i}\left( E_{\text{bound}}^{i\pm}\right)\xi_{b_i}^\dag(x)\xi_{b_i}(x),
\end{split}
\end{equation}
where the $\pm$ refers to the sign of the bound state energies, and $\xi_{b_i}=\chi_{1b_i}(\chi_{2b_i})$ for the positive (negative) bound states.
 Substituting the explicit expressions for the states given in \cite{gousheh} into Eqs.\,(\ref{r4}) and (\ref{r5}), we obtain the energy densities of the two continua and the bound states, shown bellow.
\par for $|x|\geqslant l$:
\begin{eqnarray}\label{r6}\vspace{.2cm}
\varepsilon_{\text{sky,sea}}(x)&=&\int_{0}^{+\infty}\frac{m dp}{\pi
p^2}E\bigg\{ \left(|N_{c+}|^2+|N_{c-}|^2\right)m \big[-\cos\left[2p
(|x|-l)\right]\left(|k_1|^2-p^2\right)\nonumber\\ &\times&\cosh\left[\text{Im}\left(\zeta\right)l\right]
+\sin\left[2p (|x|-l)\right]2\,p\,
\text{Im}(k_1)\sinh\left[\text{Im}\left(\zeta\right)l\right]\big ]\nonumber\\
&+&\left(|N_{c+}|^2-|N_{c-}|^2\right)\big[\cos\left[2p
(|x|-l)\right][E
\left(|k_1|^2+p^2\right)+2\,p^2\,\text{Re}(k_1)]\cos\left[\text{Re}\left(\zeta\right)l\right]\nonumber\\
&-&\sin\left[2p (|x|-l)\right]p\,
[|k_1|^2+p^2+2E\text{Re}\left(k_1\right)]\sin\left[\text{Re}\left(\zeta\right)l\right]\big]\bigg\},\\\label{r62}
\varepsilon_{\text{bound}\pm}(x)&=& 2N^2E_{\text{bound}}^{\pm}
\text{e}^{-2\lambda\left(|x|-l\right)},
\end{eqnarray}
\par and for $|x|\leqslant l$:
\begin{align}\label{r7}\vspace{.2cm}
\varepsilon_{\text{sky,sea}}(x)=&\int_{0}^{+\infty}\frac{dp}{\pi
}E\bigg\{ -1
+\left(|N_{c+}|^2+|N_{c-}|^2\right)\big[|k_1|^2+p^2+2E\text{Re}\left(k_1\right)+2m^2\big]\nonumber\\
&\times \cosh\left[\text{Im}\left(\zeta\right)x\right]
+2m\big[E+\text{Re}(k_1)\big]\cos\left[\text{Re}\left(\zeta\right)x\right]
\left(|N_{c+}|^2-|N_{c-}|^2\right)\bigg\},\\\label{r72}
\varepsilon_{\text{bound}\pm}(x)=&\frac{ 4N^2E_{\text{bound}}^{\pm}}{\left[\text{Re}\left(\zeta\right)\right]^2}
\bigg\{\left(\mu
E_{\text{bound}}^{\pm}-2\lambda^2\right)\cos\left[\text{Re}\left(\zeta\right)(x-l)\right]\nonumber\\
&-\text{Re}\left(\zeta\right)\lambda\sin\left[\text{Re}\left(\zeta\right)(x-l)\right]+\frac{1}{2}\mu^2+\mu
E_{\text{bound}}^{\pm}\bigg\},
\end{align}
where $\lambda\equiv \sqrt{m^2-E_{\text{bound}}^2}$, $\zeta \equiv \sqrt{\mu^2-4\left(\lambda^2-\mu E' \right)}$ (where $E'=\{E,E_{\text{bound}}\}$ depending on the case), $k_{1,2}\equiv \frac{1}{2}\left(\mu\pm\zeta\right)$ and the normalization constants are
\begin{align*} \vspace{.2cm}
 N&=\bigg\{\frac{2}{\lambda}+\frac{4}{\left[\text{Re}\left(\zeta \right) \right]^2}\bigg[\frac{\mu E_{\text{bound}}^{\pm} -2 \lambda^2}{\text{Re}\left(\zeta \right)}\sin\left[2 l \text{Re}\left(\zeta \right)\right]-\lambda\cos\left[2 l \text{Re}\left(\zeta \right)\right]+\lambda \nonumber\\&+ l\mu \left(\mu +2 E_{\text{bound}}^{\pm}\right)\bigg]\bigg\}^{-1/2},\nonumber\\N_{c\pm}&\equiv \bigg\{\cosh\left[\text{Im}\left(\zeta\right)l\right]\left[\left(\frac{|k_1|^2}{p^2}+1\right)2 E^2+4 E \text{Re}\left(k_1\right)\right]\mp\cos \left[\text{Re}\left(\zeta\right)l\right]2 m E\nonumber\\&\times\left(\frac{|k_1|^2}{p^2}-1\right)\bigg\}^{-1/2}.
\end{align*}
All of the spatial integrals can be performed
analytically. However, the remaining integrals over $p$ in Eqs.\,(\ref{r6},\ref{r7}) cannot be done
analytically and we integrate them numerically. Figure (\ref{fig.3}) shows the
Casimir energy as a function of $\mu$ at $\theta_0=\pi$, that is a soliton of winding number one.
This graph has a sharp maximum at the critical value $\mu\approx
2.957m$,  where a bound state crosses $E=0$ precisely at $\theta_0=\pi$. This crossing is similar to what is depicted in  Fig.\,(\ref{fig.2}) for $\mu=4m$
 where one bound state crosses $E=0$ at $\theta_0\approx0.77\pi$. The appearance of the maximum at the critical point implies that constructing
  a soliton with this slope needs
more energy than the ones with other slopes. According to Jackiw and Rebbi's
profound work \cite{jackiw}, when the system has particle conjugation symmetry and there is a zero-energy fermionic mode, the
soliton carries fermion number $\pm \frac{1}{2}$. In our problem which lacks this symmetry, our analysis shows that this point is energetically unfavorable.
 However, this property might be shared by problems possessing this symmetry.
\par One can easily obtain the behavior of the Casimir energy at $\theta_0=\pi$ numerically for large values of $\mu$ and the result is
`` $0.00163908+18.5905/x-52.6919/x^{1.25}+67.2916/x^{1.5}-47.1817/x^{1.75}+15.3126/x^2$ '', where $x=\mu/m$.
 Therefore, we can conclude that to within our numerical accuracy, it tends to
zero as $\mu\rightarrow \infty$. This is the expected result, as we
shall describe shortly. Gousheh and Mobilia obtained the
difference between the total number of levels in the Dirac sea in
the presence and absence of the soliton, and in the limit $\mu\rightarrow \infty$ their result is
as follows
\begin{equation}\label{r2}\vspace{.2cm}
Q_-=-\frac{1}{2}+\int_{0}^{+\infty}
\frac{dp}{\pi}\frac{m^2\sin(\theta_0)\cos(\theta_0)}{\sqrt{p^2+m^2}\left[p^2+m^2\sin^2(\theta_0)\right]}.
\end{equation}
Using this relation, the Casimir energy for the case $\mu\rightarrow
\infty$ would be
\begin{eqnarray}\label{r3}\vspace{.2cm}
E_{\text{Casimir}}=\frac{m}{2}-\int_{0}^{+\infty}
\frac{dp}{\pi}\frac{m^2\sin(\theta_0)\cos(\theta_0)}{\left[p^2+m^2\sin^2(\theta_0)\right]}
+\sum\limits_{i}\left(E_{\text{bound}}^{i-}\right).
\end{eqnarray}
The last term appears automatically due to our choice of the expansion of the Fermi
field in the presence of the background field (we choose the annihilation (creation) operator for positive
(negative) bound states).
When $\mu\rightarrow \infty$ one bound state just reaches the Dirac sea at $\theta_0=n \pi$, i.e. it becomes a threshold bound state.
 Therefore, like all threshold bound states in one spatial dimension,
 it should be counted as one half bound state. Since its energy is $-m$, the last term in Eq.\,(\ref{r3}) is $-m/2$.
  Therefore, the Casimir energy is zero in this limit at $\theta_0=n \pi$, since the integrand term vanishes identically there.
 For small values of $\mu$ the graph of the Casimir energy can be approximated by ``$ 0.0000252628+0.252951x-0.00272887x^{1.5}-0.015381x^2$''.
  This shows that the Casimir energy tends to zero, to within our numerical precision as $\mu\rightarrow 0$.
  Therefore, as the slope of the soliton decreases to zero, the vacuum energy approaches that of the trivial vacuum,
   in spite of the residual non-trivial boundary conditions. These two limits occur at any $\theta_0=n\pi$ i.e. when we have proper solitons with winding number $n$.
However, for arbitrary values of $\theta_0$, although the result $E_{\text{Casimir}}\rightarrow 0$ as $\mu\rightarrow 0$ still holds,
 for $\mu \rightarrow \infty$ the Casimir energy is in general non-zero and has the following form
\begin{equation}\label{r3a}\vspace{.2cm}
\begin{split}
E_{\text{Casimir}}=\frac{m}{2}\left[1+\left(1-\left[\frac{\theta_0}{\pi}\right]\right)\cos \theta_0 \right]
+m \cos \theta_0 \Theta\left(-\cos \theta_0\right) \Theta \left(\theta_0-\left[\frac{\theta_0}{\pi}\right]\pi\right),
\end{split}
\end{equation}
where $\Theta(x)$ denotes the usual step function defined by,
\begin{equation}\label{r3b}\vspace{.2cm}
\Theta(x)=\left \{ 0 \,\,\,\text{for}\,\,\, x<0 \,; \, \frac{1}{2}\,\,\, \text{for} \,\,\, x=0\,  ;\, 1 \,\,\, \text{for}\,\,\, x>0  \right\},
\end{equation}
and $\left[\frac{\theta_0}{\pi}\right]$ extracts the integer part of $\frac{\theta_0}{\pi}$.
In general the jumps in the energy of continuum and bound states at the points where threshold bound states appear cancel out so that the Casimir energy is continuous.
\par
\begin{center}
\begin{figure}[th] \hspace{3.5cm} \includegraphics[width=8.1cm]{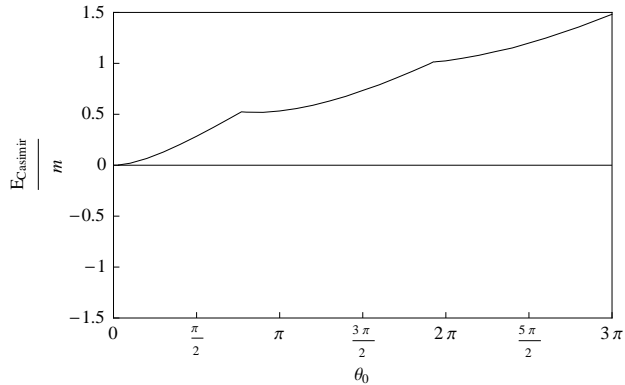}\caption{\label{fig.4} \small
  The Casimir energy as a function of $\theta_0$ for $\mu=4m$. The mild cusps in the graph are actually mild maxima and occur when bound states cross $E=0$.
   This also occurs in Fig.\,(\ref{fig.3}), except that the maximum there is much more pronounced.}
  \label{geometry}
\end{figure}
\end{center}

 In Fig.\,(\ref{fig.4}) we show the Casimir energy as a function of $\theta_0$ for $\mu=4m$.
 Note that the Casimir energy is, on the average, an increasing function of $\theta_0$ for fixed $\mu$.
Comparing Fig.\,(\ref{fig.4}) with Fig.\,(\ref{fig.2}), we can conclude that there are local maxima in the form of mild cusps in the graph of the Casimir energy when bound states cross $E=0$,
 as we vary $\theta_0$.
\par
\begin{center}
\begin{figure}[th] \hspace{.9cm} \includegraphics[width=13.cm]{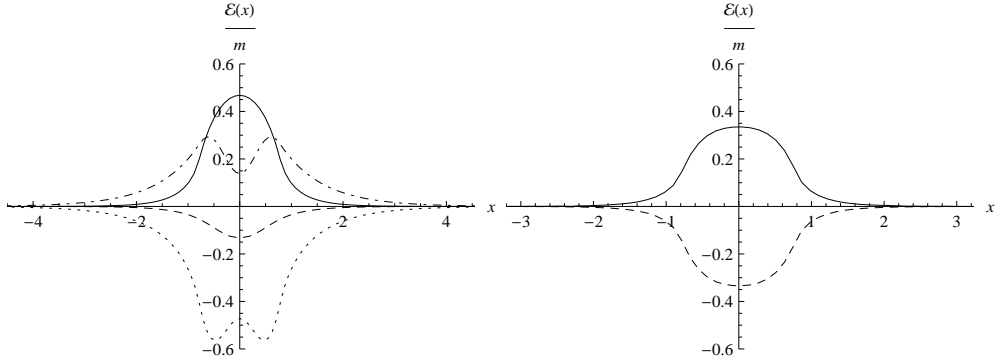}\caption{\label{fig.5} \small
  The energy densities as a function of $x$ for $\mu=4m$ at $\theta_0=\pi$. The left graph: the solid (dotted) line shows
  the energy density of the negative (positive) continuum states and dashed (dotdashed) line shows the energy density of
  the negative (positive) bound state. The right graph: the solid line shows the sum of the negative
  bound and continuum energy states and the dashed line shows the sum of the positive bound and continuum energy states. The Casimir energy density can be represented by the solid line.}
  \label{geometry}
\end{figure}
\end{center}

 In Eq.\,(\ref{e12}) the integrand of the spatial integral represents the Casimir energy density. In what follows we shall compute and display the changes for both the positive and negative parts of the spectrum for comparison.
   In general the change in the energy density naturally divides into two pieces.
    First is the contribution coming from the continuum states ($|E|\geqslant m$) and second the contribution coming from the bound states ($|E|\leqslant m$).
The left graph in Fig.\,(\ref{fig.5}) shows the energy densities of negative and positive bound and continuum states for the case $\mu=4m$ at $\theta_0=\pi$, given in Eqs.\,(\ref{r6},\ref{r62},\ref{r7},\ref{r72}).
Note that an overall positive energy density in the sea and overall negative energy density in the sky indicate that both of the continua contain spectral
deficiency as compared to the free case. The sum of these deficiencies is precisely canceled by the sum of bound state densities, at each point of space \cite{gousheh}.
 The sum of $\varepsilon_{\text{sea}}(x)$ and $\varepsilon_{\text{bound}-}(x)$ (solid and dashed lines in the left graph, respectively) is the total distortion of the states with $E<0$,
  and is depicted in the right graph in Fig.\,(\ref{fig.5}) by the solid line. Also, in this graph
   we show the sum of $\varepsilon_{\text{sky}}(x)$ and $\varepsilon_{\text{bound}+}(x)$ (dotted and dotdashed lines in the left graph, respectively), which is the total distortion of the states with $E>0$, by a dashed line.
    This figure shows that these two total densities are exactly the mirror images of each other. It is important to note that in problems which possess charge conjugation symmetry, such as ours, we can express the Casimir energy by the following equations.
 \begin{eqnarray}\label{r3b1}\vspace{.2cm}
 \varepsilon_\text{Casimir}(x)&=&\varepsilon_{\text{sea}}(x)+\varepsilon_{\text{bound}-}(x)=-(\varepsilon_{\text{sky}}(x)+\varepsilon_{\text{bound}+}(x))\\
 &=&\frac{1}{2}(\varepsilon_{\text{sea}}(x)+\varepsilon_{\text{bound}-}(x))-\frac{1}{2}(\varepsilon_{\text{sky}}(x)+\varepsilon_{\text{bound}+}(x))
 \end{eqnarray}

\par
\begin{center}
\begin{figure}[th] \hspace{3.5cm} \includegraphics[width=8.cm]{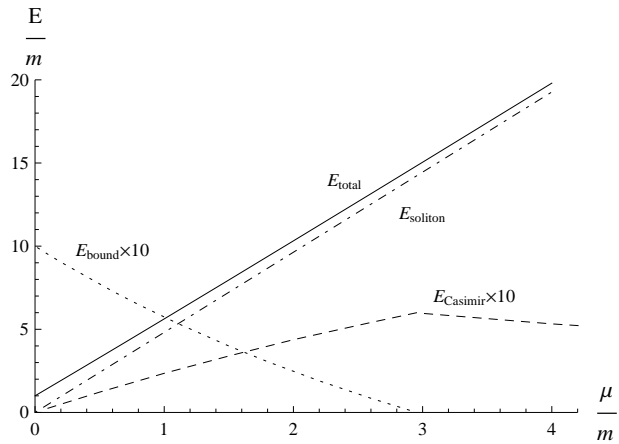}\caption{\label{fig.7} \small
  The total energy (the sum of the energy of a real fermion, the soliton mass and the Casimir energy) as a function of $\mu$.
  The solid line shows the total energy, dashed line the Casimir energy, dotdashed line the energy of the soliton ($M_{\text{cl}}$ given in Eq.\,(\ref{e1r})), and the dotted line the energy of a real fermion.
   The fermion energy and Casimir energy are multiplied by $10$ to make them more visible.}
  \label{geometry}
\end{figure}
\end{center}

 Now we can consider the effect of the Casimir energy on the total energy of the system with a real fermion present at the ground state level.
 For such a system the total energy is the sum of the soliton mass and the energy of the single fermion, if the Casimir energy is not taken into account.
 Since the ground state of the system is distorted by the presence of the soliton, this portion should be added. Note that for those values of $\mu$ for which
 the energy of the fermion is negative, we should not add this energy, since this portion has already been taken into account by the Casimir energy.
 Figure (\ref{fig.7}) shows the total energy (solid line) along with the Casimir energy (dashed line), the soliton mass (dotdashed line) and the
 energy of the fermion in the ground state (dotted line). In this figure we just depict the binding energy as long as it is positive. The main portion
  in the total energy originates from the soliton mass. Both the fermion and Casimir energies are multiplied by $10$ to make them more visible on the graph.
  
\section{Conclusion}
In this paper we have investigated a system consisting of a pseudoscalar field interacting non-linearly with a Fermi field.
 The pseudoscalar field in this model is static and prescribed, and by varying its parameters,
  that is the slope $\mu$ and its asymptotic values $\pm \theta_0$, we can form different configurations including the ones with a proper topological charge.
   Using the exact fermionic spectrum for this model, we have calculated the exact value of the Casimir energy by subtracting the vacuum energy
    in the presence and absence of the soliton. We have displayed the Casimir energy as a function of $\theta_0$ and $\mu$
    and discovered that this energy has a sharp maximum at the points where the Fermi field has zero-mode bound states
      which would correspond to a fractional fermion number for the soliton if particle conjugation symmetry was not broken.
       Therefore, our analysis shows that the configurations which have zero fermionic modes are less likely to occur.
         As expected, the Casimir energy tends to zero as $\mu\rightarrow \infty$ when $\theta_0$
          is an integer multiple of $\pi$ and as $\mu\rightarrow 0$ for all values of $\theta_0$.
           Furthermore, we have shown the Casimir energy as a function of $\theta_0$ at $\mu=4m$.
            This graph has two mild cusps at the points where a bound state crosses $E=0$, and this is the generic behavior of the Casimir energy for any values of $\mu$.
             Then, we have exhibited the contributions of the bound and continuum states to the total energy
            densities for the states with $E<0$ and $E>0$, separately and found out that they are the mirror images of each other at each point of space.
               Finally, we have added the Casimir energy to the total energy of a system consisting of a soliton and a real fermion
               in the ground state. We have concluded that although the contribution of the Casimir energy is small compared to the soliton mass, it is
               not negligible in comparison with the
               fermion energy.

\section*{Acknowledgement} We would like to thank the research office of the Shahid Beheshti University for financial support.

\end{document}